\documentclass[preprint2]{aastex6}
\usepackage{epstopdf}

\fullcollaborationName{The Friends of AASTeX Collaboration}

\shorttitle{Gas removal in \object{Ursa Minor}}
\shortauthors{Caproni et al.}

\begin{document}


\title{Gas removal in the \object{Ursa Minor} galaxy: linking hydrodynamics and chemical evolution models}


\author{Anderson Caproni\altaffilmark{1,3}, Gustavo Amaral Lanfranchi\altaffilmark{1}, Gabriel Henrique Campos Baião\altaffilmark{1}, Grzegorz Kowal\altaffilmark{1,2} and Diego Falceta-Gon\c{c}alves\altaffilmark{2}}




\altaffiltext{1}{N\'ucleo de Astrof\'\i sica Te\'orica, Universidade Cruzeiro do Sul, R. Galv\~ao Bueno 868, Liberdade, 01506-000, S\~ao Paulo, SP, Brazil}
\altaffiltext{2}{Escola de Artes, Ci\^encias e Humanidades, Universidade de S\~ao Paulo, Rua Arlindo Bettio 1000, CEP 03828-000 S\~ao Paulo, Brazil}
\altaffiltext{3}{anderson.caproni@cruzeirodosul.edu.br}

\begin{abstract}

We present results from a non-cosmological, three-dimensional hydrodynamical simulation of the gas in the dwarf spheroidal galaxy \object{Ursa Minor}. Assuming an initial baryonic-to-dark-matter ratio derived from the cosmic microwave background radiation, we evolved the galactic gas distribution over 3 Gyr, taking into account the effects of the types Ia and II supernovae. For the first time, we used in our simulation the instantaneous supernovae rates derived from a chemical evolution model applied to spectroscopic observational data of \object{Ursa Minor}. We show that the amount of gas that is lost in this process is variable with time and radius, being the highest rates observed during the initial 600 Myr in our simulation. Our results indicate that types Ia and II supernovae must be essential drivers of the gas loss in \object{Ursa Minor} galaxy (and probably in other similar dwarf galaxies), but it is ultimately the combination of galactic winds powered by these supernovae and environmental effects (e.g., ram-pressure stripping) that results in the complete removal of the gas content.

\end{abstract}




\section{Introduction} \label{sec:intro}

A puzzling feature of all dwarf spheroidal galaxies (dSph) of the Local Group is the lack of neutral gas, at least in their central regions (\citealt{mate98}, \citealt{gpu09} and references therein). \object{Ursa Minor}, as all other classical dSphs, is an example of a galaxy that does not exhibit a gaseous component. \citet{gpu09} examined the environment and HI content of all Local Group dwarf galaxies using HIPASS (\citealt{barn01}) and LAB (\citealt{kal05}) data. They did not detected any sign of neutral gas in \object{Ursa Minor}, confirming the findings of previous works (\citealt{knap78}, \citealt{will05}, \citealt{belo07}, \citealt{sige07}). These authors reached no clear conclusion on which would be the dominant physical process responsible for the disappearance of the interstellar gas, even though external mechanisms, such as ram-pressure and tidal stripping are favored (see also \citealt{eme16} and references therein). On the other hand, internal processes such as galactic winds triggered by supernovae (SNe) explosions should also be taken into account. 

Recently, \citet{ruiz13} and \citet{cap15} investigated, through 3D-hydrodynamical simulations, the role played by SNe feedback in the gas removal of dSphs and concluded that its effect depends on the SNe rate amongst other features (see also \citealt{eme16} for less massive galaxies). In such simulations, SNe occurrence is typically set numerically by a specific law (either random or deterministic), without any observational constraint. The SNe rate, however, is crucial, since it defines how much energy will be injected in the ISM and how this energy will be released, if at a constant rate through time (as in \citealt{cap15}) or constrained, e.g., by a Schmidt Law (\citealt{sch63}), as in \citet{ruiz13}. 

In this work, we provide the first 3D-hydrodynamical simulations linked to observationally constrained chemical evolution models of a galaxy. In particular, the simulations are run with the SNe (types II and Ia) rates derived by a chemical evolution model that reproduces very well several observational constraints ([$\alpha$/Fe], [Eu/Fe], [Ba/Fe] ratios, the present day gas mass, stellar metallicity distribution) of the dSph galaxy Ursa Minor (\citealt{lama04}, \citealt{lama07}). By doing that, it is assured that the SN rate is independently constrained, and therefore being more robust than other methods typically adopted in the literature.

\section{Numerical setup and initial conditions} \label{sec:NumSetup}

We used the numerical code PLUTO\footnote{\url{http://plutocode.ph.unito.it/}} \citep{mig07} to solve the classical hydrodynamic differential equations

\begin{equation} \label{massconserv}
\frac{\partial \rho}{\partial t} + \nabla\cdot\left(\rho\textbf{\emph{v}}\right) = 0,
\end{equation}

\begin{equation} \label{momconserv}
\frac{\partial \left(\rho\textbf{\emph{v}}\right)}{\partial t} + \nabla\cdot\left(\rho\textbf{\emph{v}}\textbf{\emph{v}} + P\mathbf{I}\right) = -\rho\nabla\Phi,
\end{equation}

\begin{equation} \label{energconserv}
\frac{\partial E}{\partial t} + \nabla\cdot\left[\left(E+P\right)\textbf{\emph{v}}\right] = F_\mathrm{c} - \rho\textbf{\emph{v}}\cdot\nabla\Phi,
\end{equation}
where $\rho$ is the mass density, $\textbf{\emph{v}}=(v_\mathrm{x}, v_\mathrm{y}, v_\mathrm{z})^T$ is the fluid velocity in Cartesian coordinates, $P$ is the thermal pressure,  $\mathbf{I}$ is the identity tensor of rank 3, and $E$ is the total energy density

\begin{equation} \label{totE}
E = \rho\epsilon + \frac{\rho\vert\textbf{\emph{v}}\vert^2}{2},
\end{equation}
where $\rho\epsilon$ is the internal density energy. In this work, we adopted the ideal equation of state $P = (\Gamma-1)\rho\epsilon$, where $\Gamma$ is the adiabatic index of the plasma, assumed as 5/3. This choice implies to a sound speed of the plasma, $c_\mathrm{s}$, defined as $c_\mathrm{s}=\sqrt{\Gamma P/\rho}$.

The quantity $F_\mathrm{c}$ in equation (\ref{energconserv}) is the cooling function

\begin{equation} \label{cooling}
F_\mathrm{c}=\frac{\partial P}{\partial t} = - \left(\Gamma-1\right)n^2\Lambda(T),
\end{equation}
where $n$ is the number density of the gas and $\Lambda(T)$, obtained from the interpolation of the precomputed tables provided by \citet{wie09}. To compute the values of $\Lambda(T)$, we follow \citet{cap15}, adopting [Fe/H] $\sim-2.13$, the median metallicity of \object{Ursa Minor} \citep{kir11}.

\begin{figure}
	\epsscale{1.18}
	\plotone{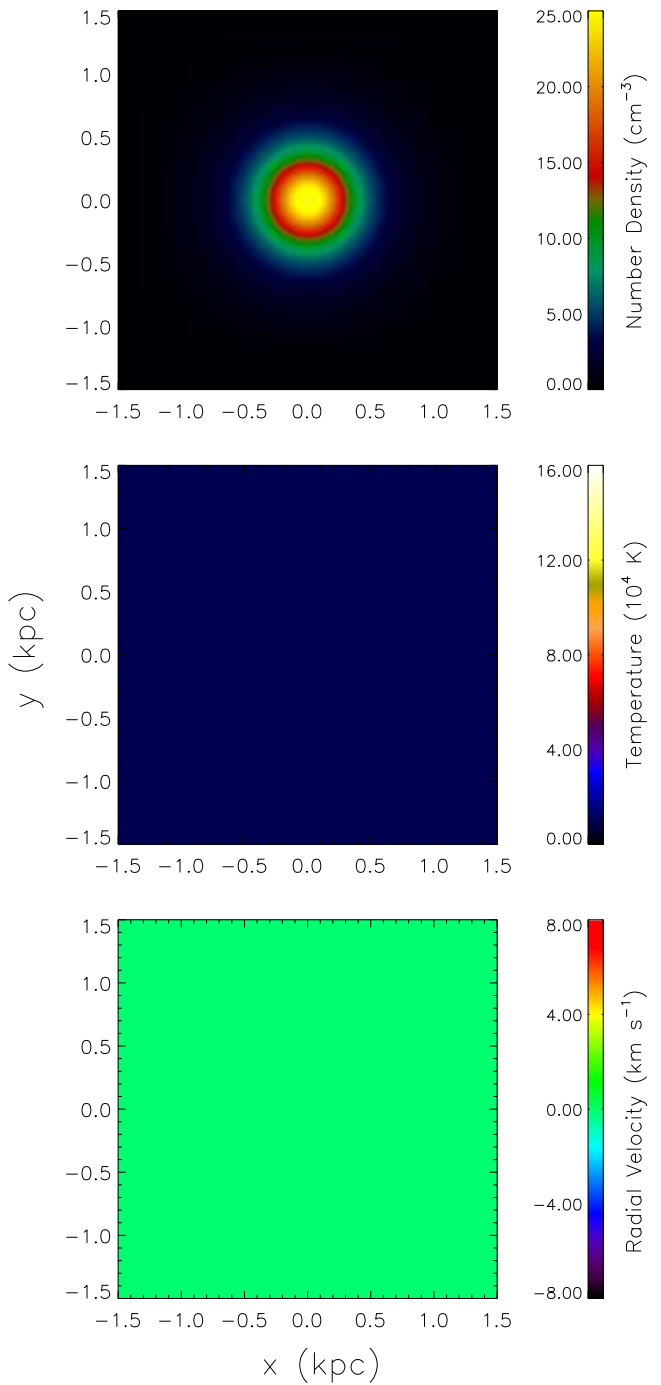}
	\caption{Initial spatial distribution of the number density ({\it top panel}), temperature ({\it middle panel}) and radial velocity ({\it bottom panel}) on $xy$ plane at $z=0$. \label{ini_profiles}}
\end{figure}

We used in our simulation the same initial gas density and pressure profiles of the models Mgh19SN1 and Mgh19SN10 found in \citet{cap15}. They were fully determined imposing hydrostatic equilibrium between an initial, isothermal gas with sound speed, $c_\mathrm{s_0}$, of 11.5 km$^{-1}$, and a cored, static dark-matter (DM) gravitational potential, $\Phi_\mathrm{h}$ \citep{bitr87,mafe99}: 

\begin{equation} \label{dm_pot} 
\Phi_\mathrm{h}(\xi) = v^2_\mathrm{c_\infty}\left[\frac{1}{2}\ln(1+\xi^2)+\frac{\arctan\xi}{\xi}\right],
\end{equation}
generated by an isothermal, spherically symmetric DM mass density profile with a characteristic radius, $r_0$, equals to 300 pc. In equation (\ref{dm_pot}), $\xi = r/r_0$, $r$ is the radial galactocentric distance, and $v_\mathrm{c_\infty}$ is the maximum circular velocity due to the DM gravitational potential, assumed to be equal to 21.1 km s$^{-1}$ (in agreement with \citealt{str07}).

 \begin{figure*}
	\plotone{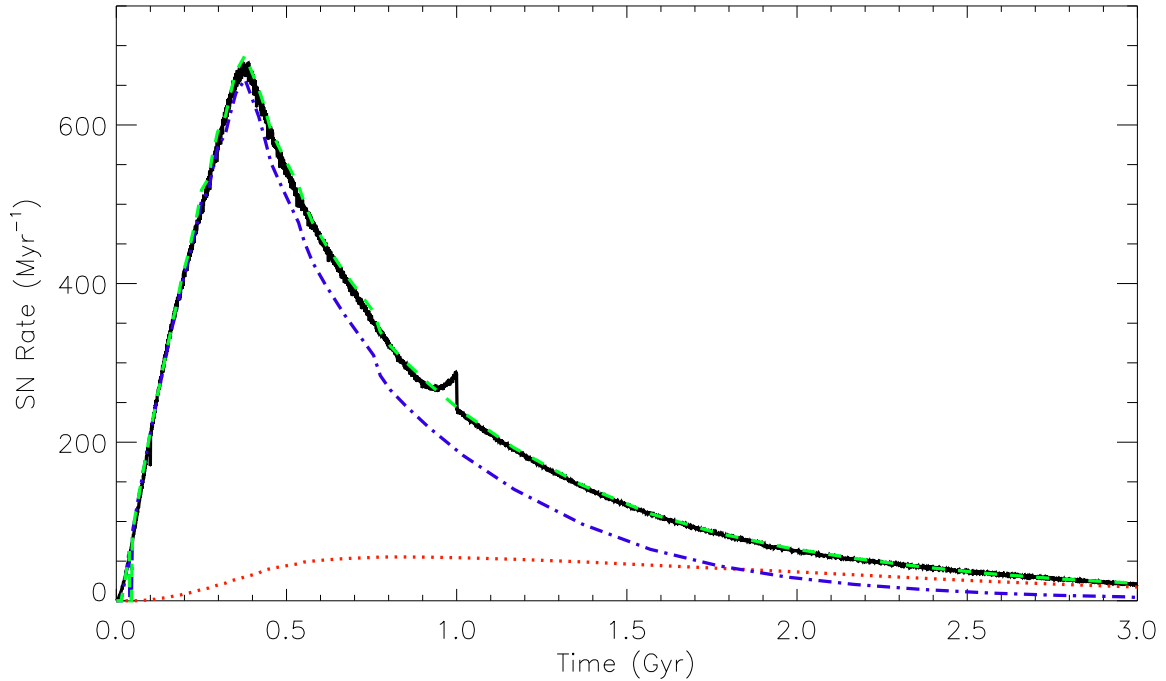}
	\caption{Supernova rates adopted in our hydrodynamical simulation of the gas in dSph galaxy \object{Ursa Minor}. Red dotted and blue dashed-dotted lines show, respectively, the instantaneous type Ia and II supernova rates derived from the chemical evolution model by \citet{lama07}. The green dashed line refers to the sum of the type Ia and II supernovae rates, while the black solid line is the total instantaneous supernovae rates found in our simulation. \label{SN_rates}}
\end{figure*}

We show in Figure \ref{ini_profiles} the initial spatial distribution of the number density, temperature, and radial velocity, $v_\mathrm{rad}$, on $xy$ plane at $z=0$. The last quantity is defined as \citep{cap15}:

\begin{equation} \label{v_rad} 
v_\mathrm{rad} = \textbf{\emph{v}}\cdot \hat{\textbf{\emph{r}}},
\end{equation}
where $\hat{\textbf{\emph{r}}}=\textbf{\emph{r}}/\vert\textbf{\emph{r}}\vert$ is the unit position vector. This initial setup implies an initial gas mass of about $2.94\times 10^8$ M$_\sun$ and a DM halo mass of $1.51\times 10^9$ M$_\sun$.

\section{Results} \label{sec:result}

As in \citet{cap15}, the Cartesian grid of the computational domain is assumed to be invariant in time, which means the effects of cosmological expansion in our simulations were neglected. We simulated a cubic region of 3 kpc $\times$ 3 kpc $\times$ 3 kpc ($\sim 1.6 R_\mathrm{{t}}$, the tidal radius of \object{Ursa Minor};  \citealt{irha95}), in which a grid with 256 points in each Cartesian direction was adopted. It translates to a numerical resolution of $\sim 11.72$ pc per cell. Equations (\ref{massconserv})--(\ref{dm_pot}) were evolved over 3 Gyr (the estimated duration of the star-formation episodes in \object{Ursa Minor}; \citealt{car02,dolp02}) through supercomputer Alphacrucis\footnote{Cluster SGI Altix ICE 8400. Further information on the Alphacrucis cluster at \url{https://lai.iag.usp.br}.} and using the message passing interface (MPI) library for parallelization\footnote{A total of about $2.03\times10^6$ processor hours were necessary to run the the simulation discussed in this work.}.

The calculation of the numerical fluxes through adjacent cell interfaces were performed from the Advection Upstream Splitting Method (AUSM+; \citealt{liou96,liou06}). The temporal evolution of $\rho$, $\rho\textbf{\emph{v}}$, and $E$ were done via the third order Runge–-Kutta method, while a piecewise total variation diminishing (TVD) linear interpolation was applied to reconstruct the primitive variables in each time step of our simulation. A closed boundary condition was assumed in the hydrodynamic calculations to minimize spurious mass fluxes into the computational domain (see Appendix A for further details).
 
Using a novel approach, we constrained the type Ia and II supernovae (SNe Ia and SNe II) rates in our simulation to those found by \citet{lama07}. We show in Figure \ref{SN_rates} these supernova rates that allow the chemical evolution model to produce [$\alpha$/Fe], [Eu/Fe], and [Ba/Fe] ratios, a stellar metallicity distribution function, and a present day gas mass in very good agreement with observations. We implemented the sum of SNe Ia and SNe II (green dashed line in Figure \ref{SN_rates}) in our simulation with a spatial distribution of energy injection modeled as follows. For the initial $1.8$ Gyr of evolution, SNe II occur at higher rates compared with the SNe Ia (Figure \ref{SN_rates}), which leads us to adopt the prescription given by \citet{cap15} (denser regions are prone to harbor an SN II event). After $1.8$ Gyr, when SNe Ia becomes dominant, the choice of SN sites becomes completely random, since the stars responsible for that type of explosion have a very long timescale (up to some gigayears), probably occurring far away from the sites where they were formed. The small deviation between theoretical and numerical SNe rates ($\la20$ SNe Myr$^{-1}$) seen between $\sim 0.9$ and 1 Gyr in the same figure is due to a small interpolation issue in the numerical implementation of the total SNe rates in our code\footnote{The total instantaneous SNe rate derived by \citet{lama07} was divided into four parts, each of them fitted by a fifth-order polynomial function to be used in our numerical code. The reported small deviation happens exactly in one of these transition epochs.}. Following \citet{fra04} and \citet{cap15}, the elected SN site receives the injection of 10$^{51}$ erg into an approximately spherical volume with a two-cell radius, $\sim$23.4 pc for the numerical resolution of our simulation (see Appendix B for the possible consequences of this particular choice on the SN feedback).

\begin{figure}
	\epsscale{1.18}
	\plotone{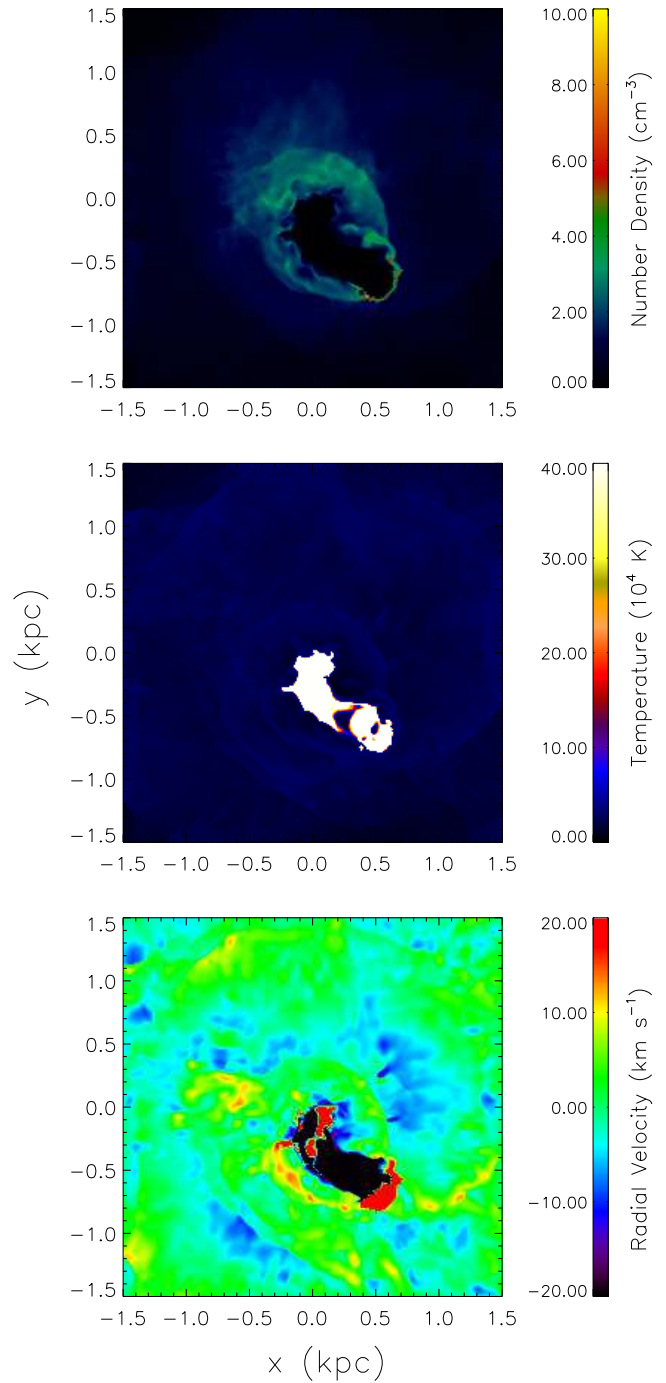}
	\caption{Spatial distribution of the number density ({\it top panel}), temperature ({\it middle panel}), and radial velocity ({\it bottom panel}) on $xy$ plane at $z=0$ after 1 Gyr of evolution. Note the extreme values of the color bar scales differ from Figure \ref{ini_profiles} by a factor of 2.5. \label{dens_temp_radv_profiles}}
\end{figure}

The dynamical impact of the implementation of SNe on the interstellar gas is noticeable. SNe perturb the initial spherically symmetric gas density distribution, as it can be seen in Figure \ref{dens_temp_radv_profiles}, driving a galactic wind responsible for removing gas from \object{Ursa Minor}. Complex structures are created by multiple interactions of the SN remnants, showing the coexistence of inflow and outflow motions (bluer and redder structures seen, respectively, in the radial velocity plot in Figure \ref{dens_temp_radv_profiles}).

\begin{figure*}
 \plotone{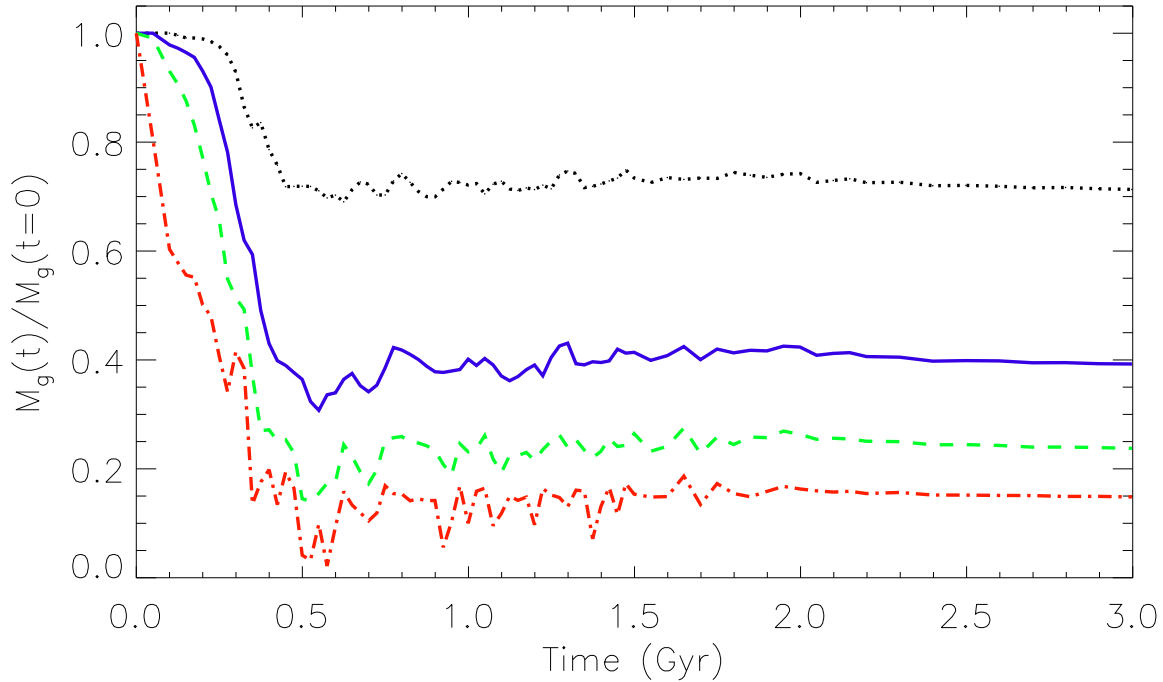}
 \caption{Instantaneous mass fraction of the gas inside a spherical concentric radii smaller than 1500 pc (black dotted line), 950 pc (blue solid line), 600 pc (green dashed line), and 300 pc (red dashed-dotted line). \label{Gas_mass_loss}}
 \end{figure*}

The successive nonlinear interactions of the SN winds push gas outward, spreading it to the outskirts of the galaxy. The total mass of the gas in the ISM of \object{Ursa Minor}, normalized by its initial value, as a function of time is displayed in Figure \ref{Gas_mass_loss}. The relative masses shown in this figure were obtained from the integration of the gas-mass density within galactocentric radii of 300, 600, 950, and 1500 pc.

An inspection of Figure \ref{Gas_mass_loss} reveals a mass loss that varies in time, as well as in galactic radius during 3 Gyr of evolution. The more accentuated losses occurred approximately in the first 600 Myr for all radii. After this epoch, gas losses slowed down substantially. This behavior agrees with the type II SN rate adopted in this work, which peaks roughly around 400 Myr. In terms of radius, the central region of \object{Ursa Minor} ($r\le300$ pc) lost most of its gas, as high as 90\% of the initial mass between 500 and 600 Myr. The gas mass decreased in $\sim$80\% at the same period for the volume delimited by a radius of 600 pc\footnote{The majority of stars in \object{Ursa Minor} have been detected inside a galactocentric radius of 600 pc (e.g., \citealt{irha95})}. Inside the adopted tidal radius of \object{Ursa Minor}, 950 pc, SNe removed a fraction of $\sim$60\% in gas.

We found a higher gas loss rate in comparison with the models Mgh19SN1 and Mgh19SN10 of \citet{cap15}\footnote{Models Mgh19SN1 and Mgh19SN10 adopted, respectively, constant SN rates of 1 SN Myr$^{-1}$ and 10 SNe Myr$^{-1}$.} (e.g., a factor of $\sim 2.3$ in the case of model Mgh19SN10 - the one that has the highest gas loss rate), in which the initial gas configuration is the same as that adopted in this work. It is a straightforward consequence of a total number of SNe in our simulation larger than those assumed in \citet{cap15}. 

At the end of our simulation, the gas temperature is between 10$^4$ and 10$^5$ K, with very low cooling rates because of low gas density. About 90\% of the gas inside the computational domain has a number density between 0.1 and 1.0 cm$^{-3}$. At these physical conditions, it is expected that hydrogen atoms are fully ionized. Assuming a gas composed of pure hydrogen, we obtain a \ion{H}{2} mass of $\sim4.17\times 10^7$ M$_\sun$ inside 600 pc and after 3 Gyr of evolution. This is roughly two orders of magnitude higher than the upper limit of $\sim 10^5$ M$_\sun$ for \ion{H}{2} mass at the present time determined by \citet{gal03}.

The resulting gas mass higher than the inferred value through observations is similar to those found by \citet{cap15} but using time-constant (type II) SN rates. They pointed out a possible additional external mechanism to reconcile the final gas mass derived from their hydrodynamical simulations and that inferred observationally, such as tidal stripping (e.g., \citealt{blro00,rea06a}) and ram-pressure stripping (e.g., \citealt{gugo72,may06,mcc07}). Another possibility is whether cosmic rays or radiative heating of the local ISM increase the gas temperature. Thermal galactic winds have been shown to be responsible for relatively large mass-loss rates, though smaller than those found from SNe feedback \citep{falc13}. These possibilities cannot be ruled out by our present simulation and deserve to be investigated in future works.

It is important to emphasize that our simulation was halted at 3 Gyr, leaving more than 10 Gyr of gas evolution in \object{Ursa Minor}  unexplored. Indeed, SNe Ia events must occur after 3 Gyr, but at a lower rate ($\la 15$ SNe Myr$^{-1}$; \citealt{lama04,lama07}). Thus, SNe Ia could maintain the galactic wind active, removing an additional amount of gas from \object{Ursa Minor}. We made a simple estimate of the gas-mass fraction inside galactocentric radii of 600 and 950 pc after 13 Gyr of evolution. It was made from the extrapolation of a least-squares exponential fit of the instantaneous mass fraction as a function of the instantaneous type Ia SNe rates from the chemical evolution model by \citet{lama04} between 1.8 and 3 Gyr. We obtained mass fractions of $\sim 0.37$ and $\sim 0.22$ inside 950 pc and 600 pc, respectively. It corresponds to a decrement of less than 2\% in relation to the respective values at 3 Gyr, insufficient to substantially reduce the remaining gas mass in the end of our present simulation. 

An additional possibility to minimize the differences in the gas masses is decreasing the initial gas mass in the simulation, as raised by \citet{cap15}. They showed that a decrement of about a factor of four in the gas-to-DM mass ratio increased the fractional gas loss by a factor of $\sim 1.5$. Even though a decrease of the initial gas mass by this factor can reduce the differences between numerical and observational results, it seems insufficient to reconcile the final remaining mass in our simulation with the upper limit for the \ion{H}{2} mass in \object{Ursa Minor} nowadays.

\section{Discussion} \label{sec:concl}

For the initial thermodynamical conditions of the gas and SN rate history assumed in this work, our numerical results indicate that SN feedback plays an important role in the gas loss, removing $\sim$60\% of the original gas content inside the tidal radius of \object{Ursa Minor}. However, SN feedback was not able to completely remove the baryonic matter from \object{Ursa Minor}. This finding strengthens the results of previous works on gas losses in dwarf galaxies. Using analytic arguments and cosmological hydrodynamical simulations, \citet{rea06b} found that SN feedback is important for dwarf galaxies to keep gas hot and spread, not necessarily for ejecting all the gas to intergalactic medium (IGM). \citet{mafe99} derived a relationship between the mass of gas and the mechanical luminosity associated with SN feedback. For a gas mass between $\sim 9\times10^7$ and $3\times10^8$ M$_\sun$, and mechanical luminosities between $\sim10^{38}$ and $10^{40}$ erg s$^{-1}$ (extreme values found in our simulation), the blowout case is expected, allowing partial removal of the gas by SN feedback (see Figure 1 in \citealt{mafe99}).

From a reanalysis of the Sloan Digital Sky Survey Data Release 8, \citet{geh12} found no star-forming dwarf galaxies with stellar masses between $10^7$ and $10^9$ M$_\sun$ at distances larger than about 1.5 Mpc from the host (massive) galaxy. They also showed that the fraction of star-forming quenched dwarf galaxies increases strongly with decreasing distance from a massive host galaxy (see their figure 6). These findings imply that those field dwarf galaxies could not stop their star formation by themselves (in the sense of a passive evolution converting gas into stars), needing some extra mechanism (probably ram-pressure and tidal stripping effects) to remove gas. In this context, \object{Ursa Minor} can be classified as a non-field, quenched galaxy nowadays, since it is located at a heliocentric distance of about 64 kpc \citep{irha95}, it is not currently forming stars (e.g., \citealt{dolp02,lama04}), and it is a gas depleted system (e.g., \citealt{gpu09,spe14}). However, it is important to emphasize that \object{Ursa Minor} has an inferred stellar mass of about $8\times 10^5$ M$_\sun$ \citep{dewo03,orb08}, roughly an order of magnitude below the stellar mass range in which the results in \citet{geh12} are strictly valid.

Using a statistical approach to study the DM subhaloes produced in the Via Lactea II cosmological simulation \citep{die08}, \citet{roc12} estimated
the individual infall times of the Milky Way dwarf satellites. Defining infall time as the look-back time since the DM subhalo last crossed inward through the virial radius of the main DM halo, \citet{roc12} found that it must have happened $\sim8-11$ Gyr ago in the case of \object{Ursa Minor}. Similar to the other classical dSph galaxies, this infall time coincides roughly with the end of star-formation activity in \object{Ursa Minor} (see figure 7 in \citealt{roc12}). It suggests the motion of \object{Ursa Minor} into the gaseous halo of Milky Way could have removed gas from this satellite.

How do our results relate to these previous works? Our HD simulation showed that the adoption of type Ia and II SN rates derived from chemical evolution model by \citet{lama07} produced strong SN feedback during the first 600 Myr of evolution. At this time, \object{Ursa Minor} still had not crossed the virial radius of the Milky Way considering the infall time estimated by \citet{roc12}. Note that $\sim 60\%$ of the initial gas mass inside the tidal radius was pushed outward by those SNe (see Figure \ref{Gas_mass_loss}), decreasing substantially the average density of the gas. After 600 Myr, the type Ia and II SN rates are already decreasing in time (see Figure \ref{SN_rates}), so that SN feedback is not able to remove much more gas, just keeping it spread and heated (as found in \citealt{rea06b}). As the gas is more sparse than its initial configuration, the efficiency of ram-pressure stripping in removing gas is substantially increased. As SN feedback removes more than half of the gas after only 600 Myr, the ram-pressure mechanism can act efficiently upon \object{Ursa Minor}, blowing away the remaining gas during the rest of \object{Ursa Minor}'s life, transforming this galaxy into a quenched system (in agreement with \citealt{geh12}). This possibility might reconcile the amount of bounded gas in our simulation with the current upper limit derived observationally by \citet{gal03}. An additional simulation incorporating the ram-pressure stripping mechanism should be pursued to corroborate this scenario.

\section{Conclusions} \label{sec:concl}

Adopting a novel approach, in which instantaneous rates of the types II and Ia SN were constrained to the chemical evolution model by \citet{lama04}, we performed a three-dimensional hydrodynamical  simulation of the gas in the dSph galaxy \object{Ursa Minor} over 3 Gyr of evolution. Our main results are summarized as follows.

\begin{itemize}

 \item Mass loss is not constant in time, as well as per galactic radius, in agreement with \citet{cap15};
 
 \item The more pronounced losses occurred approximately in the first 600 Myr for all radii, in which type II SN rate adopted in this work reaches its maximum. The central region of \object{Ursa Minor} ($r\le300$ pc) lost the largest amount of gas ($\sim 90\%$ of the initial mass) in the same interval;
 
 \item After 3 Gyr of evolution, the remaining mass in gas inside 300 pc from the galaxy center is only $\sim 15\%$ of its original value. For $r\le 600$ pc, $\sim 24\%$ of the initial mass still remains, increasing to $\sim 39\%$ inside the tidal radius of \object{Ursa Minor}. Note that our closed boundary conditions tend to retain gas inside the computational domain. In addition, over cooling issues concerning the injection of SN energy into cells with number densities larger than 1 cm$^{-3}$ decreases the dynamical impact of SNe to drive gas outward (see appendices A and B, respectively). These facts imply that our estimates of gas-mass losses presented in Figure \ref{Gas_mass_loss} are slightly underestimated, reinforcing the importance of the SN mechanical feedback on the gas removal in objects similar to the \object{Ursa Minor} galaxy. These findings indicate that type II and Ia SNe explosions are probably one of the main mechanisms behind the gas losses in \object{Ursa Minor}, as well in other similar dwarf galaxies;
 
 \item \citet{gal03} derived an upper limit of $\sim 10^5$ M$_\sun$ for \ion{H}{2} mass in \object{Ursa Minor} at the present time, which is roughly two orders of magnitude lower than the mass left inside a galactocentric radius of 600 pc at the end of our simulation. The decrease of the initial gas mass and/or the extrapolation of the effects of type Ia SNe for more than 10 Gyr of evolution could diminish the discrepancy between the remaining mass in our simulation and the nowadays inferred upper limit of \ion{H}{2} mass. However, additional external mechanisms, such as tidal and/or ram-pressure stripping, must play a role in the gas loss history of \object{Ursa Minor}.
 
 \item Our numerical simulation shows that SN feedback was able to push more than half of the initial amount of gas to outer radii during the first 600 Myr of evolution. This means SNe driven galactic winds lowered the average gas densities very early in the evolutionary history of the \object{Ursa Minor} galaxy, which is crucial to make environmental effects (e.g. ram-pressure stripping) very effective at removing this gas upon infall. We believe this scenario may also reconcile the two orders of magnitude disagreement between the final gas mass found in our simulation and the current upper limit for \ion{H}{2} mass in \object{Ursa Minor}. Thus, our simulation suggests types Ia and II SN feedback is an essential player behind gas loss in \object{Ursa Minor} (and probably in other similar dwarf galaxies), but it is ultimately the combination of galactic winds powered by SNe and subsequent environmental effects that seems to remove all of the gas.
 
\end{itemize}

\acknowledgments

This work has made use of the computing facilities of the Laboratory of Astroinformatics (IAG/USP, NAT/UCS), whose purchase was made possible by the Brazilian agency FAPESP (grant 2009/54006-4) and the INCT-A. A.C. thanks the Brazilian agencies CNPq (grant 305990/2015-2) and FAPESP (grant 2017/03173-4). G.A.L. thanks the Brazilian agency CNPq (grant 304928/2015-1). D.F.G. thanks the Brazilian agencies CNPq (grant 302949/2014-3) and FAPESP (grant 2013/10559-5) for financial support. G.K. acknowledges support from FAPESP (grant 2013/04073-2) and CAPES (PNPD 1475088). The authors acknowledge financial support from FAPESP through the grant 2014/11156-4. The authors also thank the anonymous referee for a very constructive report that substantially improved the presentation of this work.







\appendix


\section{Boundary Conditions and Their Influence on the Gas-mass loss}

	In our simulation, we used closed boundary conditions, which means imposing a zero-value for the velocity component perpendicular to the computational frontiers, $v_\perp$, and zero-gradient for the other velocity components as well for pressure and mass density quantities. The standard open boundary condition (zero-gradient for all primitive variables) was found to be inappropriated for our numerical studies since it introduced too much spurious mass into the computational domain after 3 Gyr of passive evolution ($\sim 10^{13}$ M$_\sun$ inside a spherical radius of 950 pc, against $\sim 10^{6}$ M$_\sun$ using our closed boundary condition; \citealt{cap15}).
	
	Although our closed boundary condition minimizes spurious gas accretion, it can induce reverse shocks when the gas pushed by SN blasts reaches the computational frontiers, which could decrease the SN-driven gas losses. The SN-feedback-driven winds reach the edge of our simulation box for the first time after an elapsed time of about 125 Myr, and with subsonic speeds of $\sim 2$ km s$^{-1}$. To estimate the impact of the boundary condition on mass-loss rates, we calculated the gas fraction that must have been lost comparing the gas velocity in the adjacent cells at the boundaries with escape velocity, $v_\mathrm{esc}$, defined as \citep{bitr87}
	
	\begin{equation} \label{esc_veloc}
	v_\mathrm{esc}=\sqrt{2\Phi_\mathrm{h}(\xi_\mathrm{h})},
	\end{equation}
	where $\xi_\mathrm{h} = r_\mathrm{h}/r_0=49.96$ for our simulation, and $r_\mathrm{h}$ is the radius of the DM halo.
	
	If the conditions $v_\perp\ga v_\mathrm{esc}$ and $\textbf{\emph{v}}\cdot \hat{\textbf{\emph{r}}} > 0$ are simultaneously satisfied in a given adjacent cell at the computational frontiers, the gas inside it is considered unbounded to the DM halo, implying that it will not fall back into the galaxy\footnote{We neglected any deceleration that the unbounded gas could suffer from interactions with the IGM outside the boundaries.}.
	
	Let $f_\mathrm{esc}(t_i)$ be the number of unbounded cells at a time $t_i$, divided by the total number of adjacent cells to the boundaries. We can estimate the fraction of gas that must be lost due to the escape velocity condition, $f_\mathrm{g,lost}$ at a time $t_i$ through
	
	\begin{equation} \label{fgas_lost}
	f_\mathrm{g,lost}(t_i)=f_\mathrm{esc}(t_i)+f_\mathrm{g,lost}(t_{i-1})\left[1-f_\mathrm{esc}(t_i)\right],
	\end{equation}
	where $t_i>t_{i-1}$, $f_\mathrm{g,lost}(t_{i-1})=1-M_\mathrm{g}(t_{i-1})/M_\mathrm{g}(0)$, and $M_\mathrm{g}(t)$ is the instantaneous total mass of gas inside the whole computational domain.
	
\begin{figure*}
		\plotone{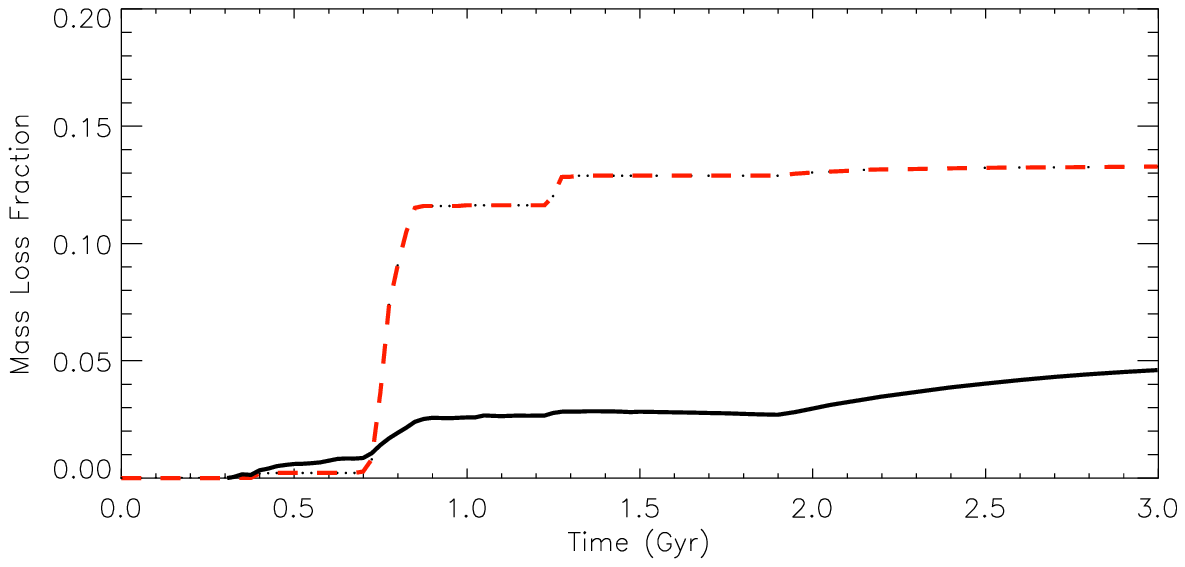}
		\caption{Time behavior of the instantaneous gas mass normalized by the initial total mass inside the whole computational domain that was lost during 3 Gyr of evolution. The black solid line refers to this fraction derived from our simulation, while the red dashed curve shows the estimation of the gas fraction lost by the galaxy considering the criteria related to escape velocity from the DM halo.  \label{M_loss_frac_vesc}}
	\end{figure*}

	We show in Figure \ref{M_loss_frac_vesc} the time behavior of $f_\mathrm{g,lost}$ during 3 Gyr of evolution (red dashed line). The same figure also displays the instantaneous normalized mass-loss fraction derived from our numerical simulation considering the whole computational domain (black solid line). Comparing both curves, we note that $f_\mathrm{g,lost}$ predicts gas losses higher than those found in our simulation (a factor of about 2.5 at 3 Gyr), suggesting that our closed boundary conditions tend to retain gas inside the computational domain. These results imply that our estimates of gas losses presented in Figure \ref{Gas_mass_loss} are slightly underestimated, reinforcing the important role of the SN mechanical feedback on the gas removal in objects similar to the \object{Ursa Minor} galaxy.
	
\section{Time evolution of a supernova remnant: numerical tests}

\begin{figure*}
	\plotone{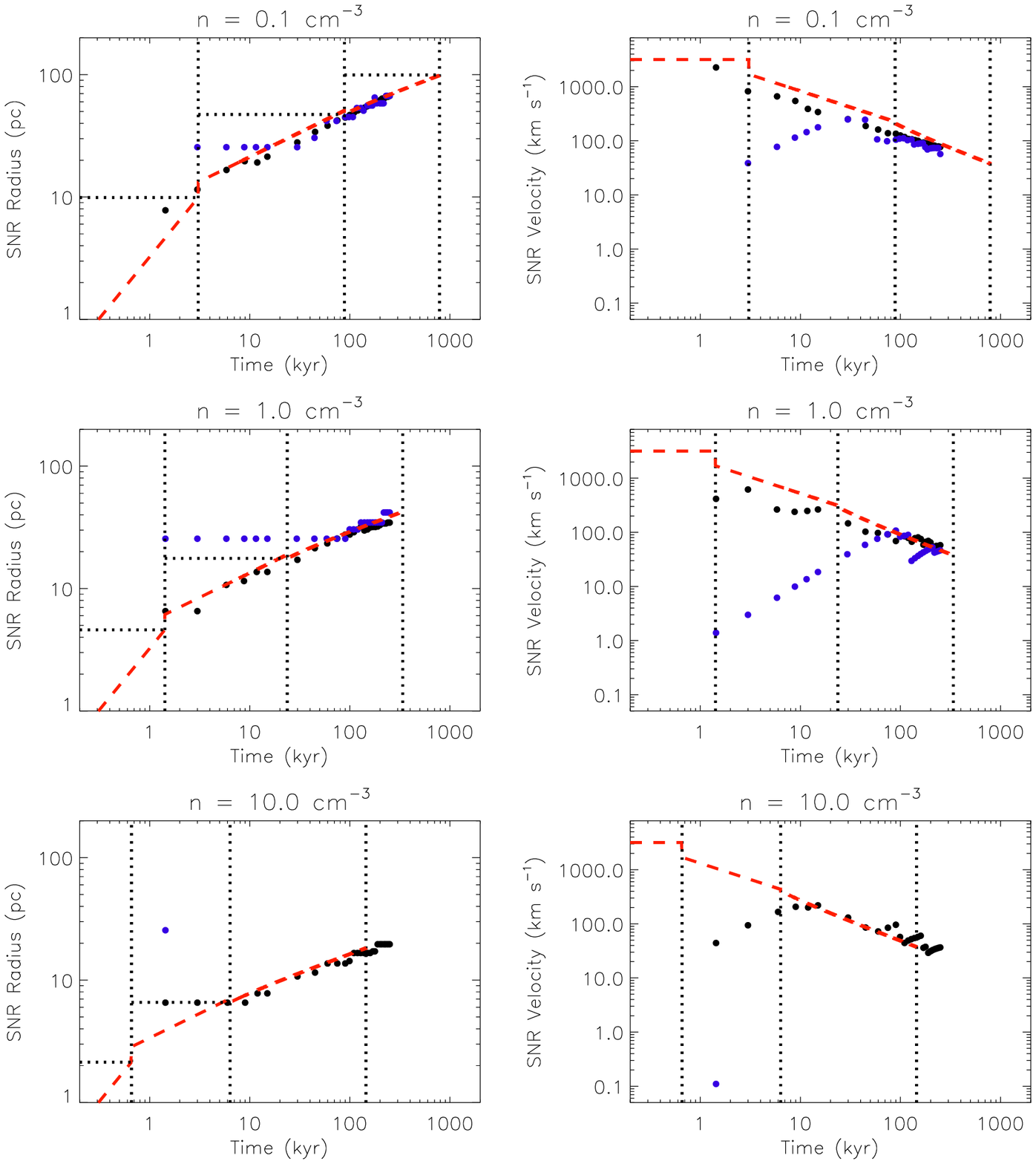}
	\caption{Time evolution of the SNR radius (left panels) and velocity (right panels) for two different numerical resolutions: 3.0 pc per cell (black circles) and 11.71875 pc per cell (blue circles). From top to bottom, ISM density is increased by a factor of 10 from 0.1 to 10.0 cm$^{-3}$. Red dashed lines represent predictions from the analytical model by \citet{cio88}. Dotted horizontal and vertical lines mark, respectively, approximated transition radii and transition epochs for the different main phases of a SNR (blast wave, Sedov--Taylor, snowplow, and merge) based on \citet{cio88}.  \label{SNR_tests}}
\end{figure*}

We present in this section validation tests concerning the time evolution of an isolated supernova remnant, during 0.25 Myr, through a homogeneous ISM with an isothermal temperature of $10^5$ K. We run HD simulations for three different ISM number densities typically found in our simulation for the \object{Ursa Minor} galaxy: 0.1, 1.0, and 10.0 cm$^{-3}$. The spatial resolution adopted in those tests was $\sim$11.72 pc per numerical cell, the same numerical resolution of our simulation presented in this paper, and 3.0 pc per cell. It allowed us to check the influence of numerical resolution on the shock evolution driven by a supernova. Those validation tests are compared to the analytical model by \citet{cio88}, as can be seen in Figure \ref{SNR_tests}.

The injection of SN energy occurs at $t=0$ and at the center of the computational domain, following the prescription adopted in \citet{fra04} and \citet{cap15}, as well as in this present paper: injection of 10$^{51}$ erg into an approximately spherical volume with a two-cell radius (6 pc for simulations with resolution of 3 pc per cell and $\sim$23.4 pc for a resolution of $\sim$11.72 pc). This increment in pressure produces an initial supersonic expanding shock with an instantaneous radius, $R_\mathrm{SNR}$, and speed, $v_\mathrm{SNR}$. 

For $n\la 1$ cm$^{-3}$, Figure \ref{SNR_tests} shows that $R_\mathrm{SNR}$ is compatible with the analytical model by \citet{cio88} in the snowplow phase ($t \ga 0.1$ Myr for $n\sim 0.1$ $^{-3}$ and $t \ga 0.02$ Myr for $n\sim 1.0$ cm$^{-3}$), independent of the numerical resolution adopted in those tests. The same conclusion is valid for $v_\mathrm{SNR}$. On the other hand, only the high-resolution test (3 pc per cell) produces $R_\mathrm{SNR}$ and $v_\mathrm{SNR}$ compatible with theoretical expectations for $n = 10.0$ cm$^{-3}$. Indeed, no shock is detected in the snowplow phase for a resolution of $\sim$11.72 pc, indicating possible over cooling issues in this case (e.g, \citealt{cre11,sim15}). This implies that the kinetic energy/momentum transferred to the shocked ISM is lower than what would be expected, weakening the kinetic feedback from a supernova for $n > 1$ cm$^{-3}$ and for a resolution of $\sim$11.72 pc per cell in our main numerical simulation.

\begin{figure*}
	\plotone{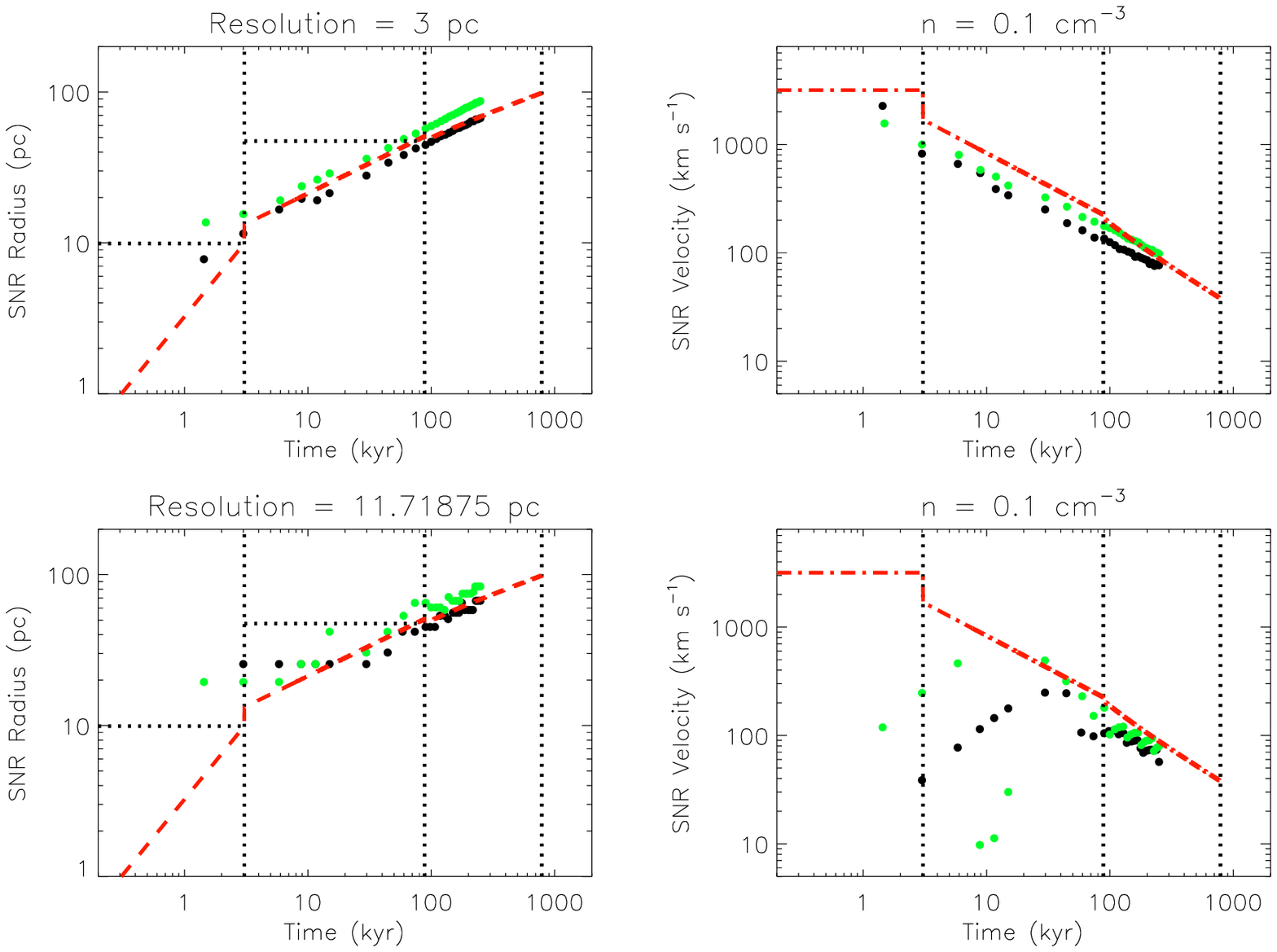}
	\caption{Time evolution of the SNR radius (left panels) and velocity (right panels) for a number density of 0.1 cm$^{-3}$ and two different numerical resolutions: 3.0 pc per cell (top panels) and 11.71875 pc per cell (bottom panels). Green and black circles refer, respectively, to SN energy injection radii of one and two numerical cells. Red dashed lines represent predictions from the analytical model by \citet{cio88}. Dotted horizontal and vertical lines mark, respectively, approximated transition radii and transition epochs for the different main phases of a SNR (blast wave, Sedov--Taylor, snowplow, and merge) based on \citet{cio88}.  \label{SNR_tests_2}}
\end{figure*}

We also checked the influence of injecting SN energy into a two-cells spherical region. With this objective, we run two additional simulations considering an injection radius of one numerical cell, $n = 0.1$ cm$^{-3}$, and numerical resolutions of 3 and $\sim$11.72 pc per cell. The results are shown in Figure \ref{SNR_tests_2}. We can realize that injecting SN energy into one cell radius increases systematically $R_\mathrm{SNR}$ and $v_\mathrm{SNR}$ in comparison with the two-cells radius case. As in \citet{cap15}, we decided to inject SN energy into a spherical radius of two cells, avoiding overestimations of the kinetic SN feedback in our HD simulations.

\end{document}